\documentclass[aoas,preprint]{imsart}

\usepackage{amsmath,amsthm,amsfonts,amsbsy,amssymb}
\usepackage{graphicx}
\usepackage{color}   
\usepackage{xcolor}   
\usepackage{bbm}
\usepackage{url} 
\usepackage{dcolumn}
\usepackage{booktabs} 
\usepackage{colortbl}
\definecolor{Gray}{gray}{0.9}
\usepackage{siunitx} 

\usepackage{multibib}
\newcites{appendix}{References}
\RequirePackage[OT1]{fontenc}
\RequirePackage{natbib}
\RequirePackage[colorlinks,citecolor=blue,urlcolor=blue]{hyperref}

\arxiv{arXiv:1711.03367} 

\startlocaldefs
\numberwithin{equation}{section}
\theoremstyle{plain}
\newtheorem{theorem}{Theorem}[section]
\newtheorem{lemma}{Lemma}[section]
\newtheorem{corollary}{Corollary}[section]

\DeclareMathOperator{\E}{\mathbb{E}}

\DeclareMathOperator{\Bias}{Bias}
\DeclareMathOperator{\V}{\mathbb{\RN{5}}}
\DeclareMathOperator{\mydiag}{diag}

\newcommand{\RN}[1]{%
  {\textup{\uppercase\expandafter{\romannumeral#1}}}%
}
\endlocaldefs

\begin{document}
\begin{frontmatter}
\title{Nonparametric Testing for Differences in Electricity Prices: The Case of the Fukushima Nuclear Accident}
\runtitle{Testing Differences in Electricity Prices}
\begin{aug}
\author{
\fnms{Dominik} 
\snm{Liebl}
\ead[label=e1]{dliebl@uni-bonn.de}}
\runauthor{D.~Liebl}
\affiliation{University of Bonn}
\address{
Dominik Liebl\\
Statistische Abteilung\\
University of Bonn\\
Adenauerallee 24-26\\
53113 Bonn, Germany\\    
\printead{e1}}
\end{aug}

\begin{abstract}
This work is motivated by the problem of testing for differences in the mean electricity prices before and after Germany's abrupt nuclear phaseout after the nuclear disaster in Fukushima Daiichi, Japan, in mid-March 2011. 
Taking into account the nature of the data and the auction design of the electricity market, we approach this problem using a Local Linear Kernel (LLK) estimator for the nonparametric mean function of sparse covariate-adjusted functional data. 
We build upon recent theoretical work on the LLK estimator and propose a two-sample test statistics using a finite sample correction to avoid size distortions.
Our nonparametric test results on the price differences point to a Simpson's paradox explaining an unexpected result recently reported in the literature.

\end{abstract}

 
\begin{keyword}
\kwd{electricity spot prices} 
\kwd{functional data analysis}
\kwd{local linear kernel estimation}
\kwd{nuclear power phaseout}
\kwd{sparse functional data}
\kwd{time series analysis}
\end{keyword}

\end{frontmatter}

\section{Introduction}\label{sec:intro}
On March 15, 2011, Germany showed an abrupt reaction to the nuclear disaster in Fukushima Daiichi, Japan, and shut down $40\%$ of its nuclear power plants---permanently. This substantial loss of cheap (in terms of marginal costs) nuclear power raised concerns about increases in electricity prices and subsequent problems for industry and households. So far, however, empirical studies are scarce and based on restrictive model assumptions. In this work we add a nonparametric functional data perspective and compare our test results with the existing benchmark results. Our results point to a Simpson's paradox explaining the unexpected result recently reported by \cite{GHW2017}.

Pricing at electricity exchanges is explained well by the  merit-order model. This model assumes that spot prices are based on the merit-order curve---a monotonically increasing curve reflecting the increasingly ordered generation costs of the installed power plants. The merit-order model is a fundamental market model \citep[see, for instance,][Ch.~4]{Burger2008} and is most important for the explanation of electricity spot prices in the literature on energy economics \citep[see][]{burger2004spot, sensfuss2008merit, hirth2013market, L13, cludius2014merit, BKF17, GHW2017}.

\begin{figure}[t]
\centering  
\includegraphics[width=\textwidth]{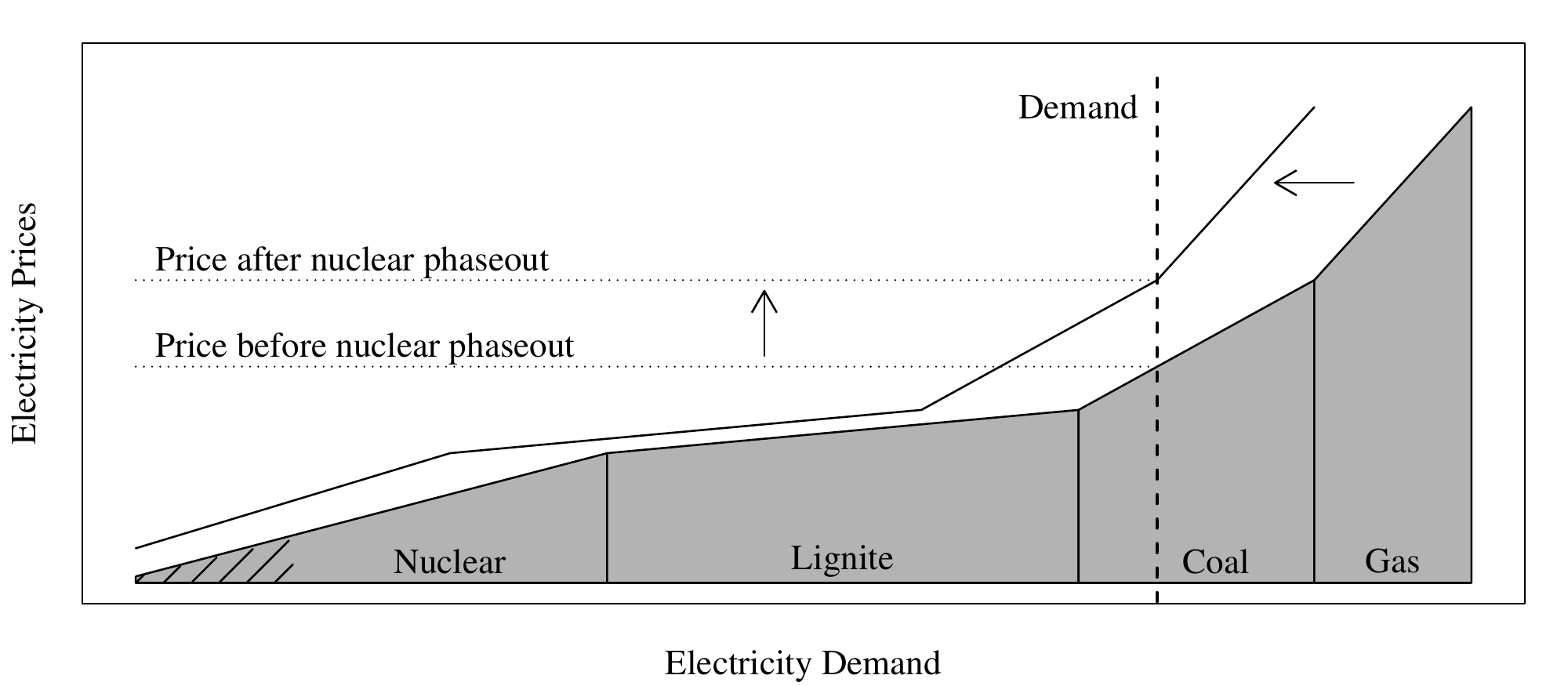}
\caption[]{Sketch of the merit-order curve and the theoretical price effect of the nuclear power phaseout. The dashed region signifies the proportion of phased out nuclear power plants.}
\label{Fig:MOC}
\end{figure}

The plot in Figure \ref{Fig:MOC} sketches the merit-order curve of the German electricity market and is in line with \cite{cludius2014merit}. The interplay of the demand curve (dashed line) with the merit-order curve determines the electricity prices, where electricity demand is assumed to be price-inelastic in the short-term perspective of a spot market. The latter assumption is regularly found in the literature \citep[see, e.g.,][]{sensfuss2008merit} and confirmed in empirical studies \citep[see, e.g.,][]{Lijesen2007249}. 

We consider electricity spot prices from the European Power Exchange (EPEX), where the hourly electricity spot prices of day $i$ are settled simultaneously at $12$ am the day before \citep[see, for instance,][Ch.~6]{BKL13}. Following the literature, we differentiate between ``peak-hours'' (from 9am to 8pm) and ``off-peak-hours'' (all other hours) and focus on the $m=12$ peak-hours, since these show the largest variations in electricity prices and electricity demand.

\begin{figure}[htb]
\centering
\includegraphics[width=1\textwidth]{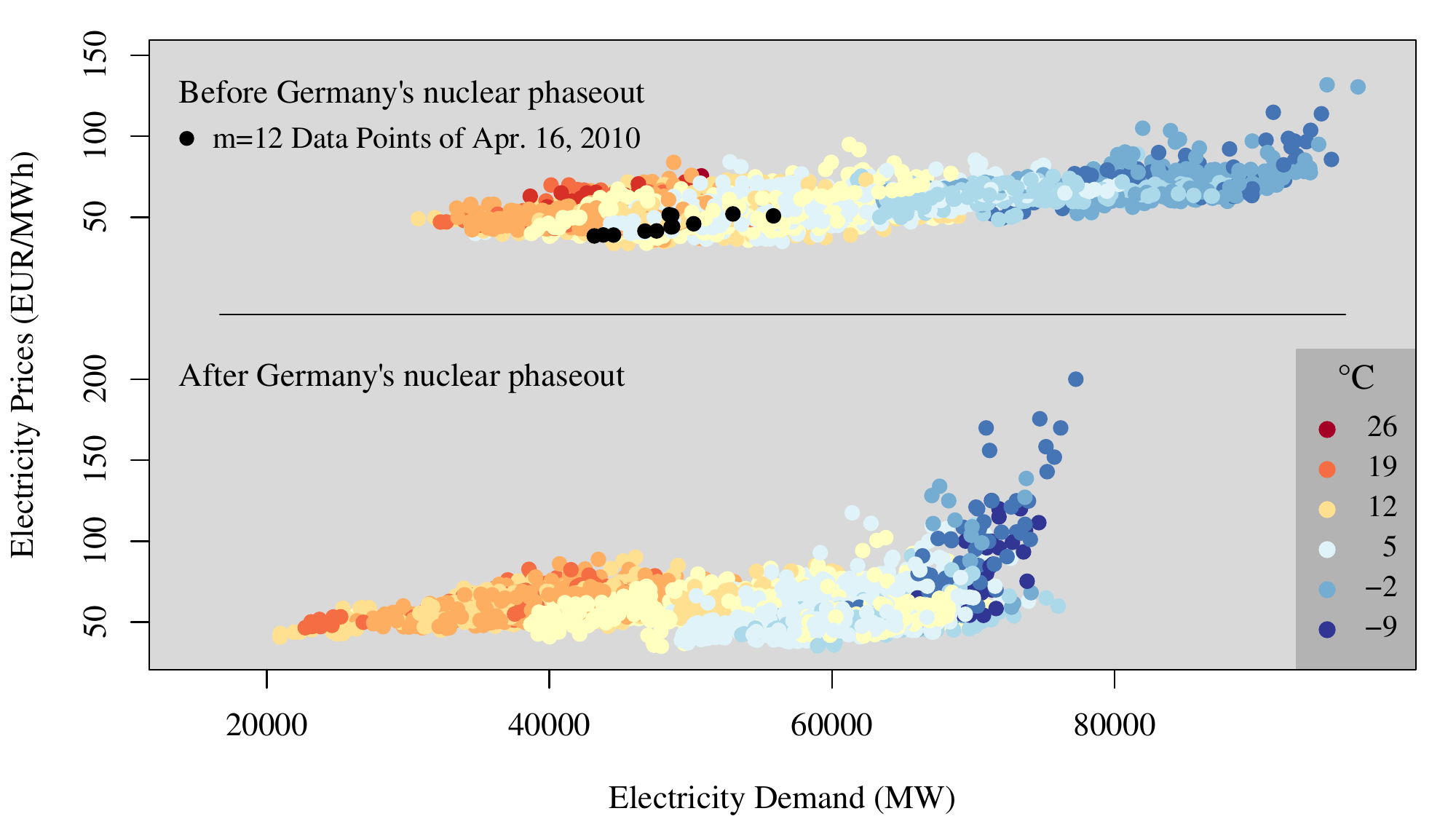} 
\caption[]{Scatter plots of the price-demand data pairs $(Y_{ij},U_{ij})$ and the the additional covariate of daily mean air temperature $Z_i$ (measured in $^\circ C$). The upper panel shows the data from one year before Germany's partial nuclear phaseout, i.e., from March 15, 2010 to March 14, 2011; the lower panel shows the data from one year after, i.e., from March 15, 2011 to March 14, 2012.}
\label{Fig:SP} 
\end{figure}

The daily simultaneous pricing scheme at the EPEX results in a daily varying merit-order curve (or simply ``price curve'') $X_i$. However, we do not directly observe the price functions $X_i$, but only their noisy discretization points $(Y_{ij},U_{ij})$, with $j=1,\dots,m=12$ (see black points in Figure \ref{Fig:SP}). This data situation with only a few, i.e., $m=12$, irregularly spaced evaluation points, $U_{i1},\dots,U_{im}$, per function is referred to as sparse functional data \citep[see, e.g.,][]{Yao2005}. The smoothness of the underlying price curve $X_i$ induces a high correlation between electricity prices $Y_{ij}$ and $Y_{ik}$ with similar values of electricity demand $U_{ij}\approx U_{ik}$. Ignoring these correlations when doing inference can result in serious size distortions and invalid test decisions \citep[see][]{DL2018}.

Therefore, we model the electricity spot price $Y_{ij}\in\mathbb{R}$ of day $i$ and peak-hour $j$ as a discretization point of the underlying (unobserved) daily merit-order curve $X_i$ evaluated at the corresponding value of electricity demand $U_{ij}\in\mathbb{R}$, 
\begin{align}\label{eq:basic_1}
  Y_{ij}&=X_i(U_{ij},Z_i)+\epsilon_{ij},\quad j=1,\dots,m,\quad i=1,\dots,n,
\end{align}
where $Z_i\in\mathbb{R}$ denotes the daily mean air temperature---a covariate of fundamental importance for the shape and the location of the random function $X_i(.,Z_i)$. The statistical error term $\epsilon_{ij}$ is assumed to be independently and identically distributed (iid) with mean zero and finite variance and assumed to be independent from $X_i$, $U_{ij}$, and $Z_i$. 

The shape of $X_i(.,Z_i)$ and its location, i.e., $X_i(.,Z_i)\in L^2[a(Z_i),b(Z_i)]$, with $[a(Z_i),b(Z_i)]\subset\mathbb{R}$, are both allowed to be functions of the covariate temperature $Z_i$. 
The scatter plot of the data triplets $(Y_{ij},U_{ij},Z_i)$ is shown in Figure \ref{Fig:SP}. Note that electricity demand $U_{ij}$ is observed within temperature-specific subintervals, i.e., $U_{ij}\in[a(Z_i),b(Z_i)]$ and that the discretization points $(Y_{ij},U_{ij})$ suggest steeper functions $X_i(.,Z_i)$ for cold days than for warm days. These observations motivate our modeling assumption that $X_i(.,Z_i)\in L^2[a(Z_i),b(Z_i)]$; see \cite{horvath2012inference}, Ch.~2, for fundamental properties of random functions in the space $L^2$. 

Germany's (partial) nuclear phaseout means a shift of the mean merit-order curve resulting in higher electricity spot prices---particularly at hours with large values of electricity demand (see Figure \ref{Fig:MOC}). This effect is obvious in the data for very cold days (see Figure \ref{Fig:SP}), though not obvious on other days. Therefore, the objective of this article is a two-sample test of the pointwise null hypothesis of equal means against the alternative of larger mean values after Germany's nuclear phaseout, i.e.,
\begin{align*}
H_0:\; \mu_A(u,z)=\mu_B(u,z)\quad\text{vs}\quad H_1:\; \mu_A(u,z)>\mu_B(u,z),
\end{align*}
where $\mu_A(u,z)=\E\big(X^A_i(u,z)\big)$ and $\mu_B(u,z)=\E\big(X^B_i(u,z)\big)$ are the mean functions of the random price functions $A$fter, $X_i^A(.,z)$, and $B$efore, $X_i^B(.,z)$, Germany's nuclear phaseout.

We estimate the mean functions $\mu_A$ and $\mu_B$ separately for each period $P\in\{A,B\}$ from the observed data points $\{(Y^P_{ij},U^P_{ij},Z^P_i);1\leq i\leq n_P,1\leq j\leq m\}$ using the Local Linear Kernel (LLK) estimator for sparse functional data suggested by \cite{jiang2010covariate}. Recently, it has been demonstrated that the asymptotic results of \cite{jiang2010covariate} neglect an additional functional-data-specific variance term which is asymptotically negligible, but typically not negligible in practice. Neglecting this additional variance term leads to a too small variance component resulting in size distortions of the test statistic and in invalid test decisions \citep{DL2018}. Therefore, we take into account the small sample correction proposed by \cite{DL2018} and propose a two-sample test statistic, which guarantees valid test decisions in practical finite samples as well (see also the in-depth simulation study in \cite{DL2018}).

In order to control for the effects of the remaining fundamental market factors, i.e., the price of natural gas, CO$_2$ emission allowances, and coal \citep[see, e.g.,][]{MW2016} we use what is called an event study approach. Event studies are essentially two-sample test problems comparing a ``control'' sample, from the period just before the event, with a ``treatment'' sample, from the period just after the event. The time period before the event is called ``estimation window'' and the time period after the event is called ``event window''. The idea is to choose small enough time windows for which the non-controlled, but less important, market factors do not have confounding effects \citep{McWS1997}. The event study method used today was introduced by \cite{BB1968} and \cite{FFJR1969}. A well-known introductory survey article is written by \cite{MacK1997}. We are not the first to use the event study approach for analyzing effects that are due to Germany's unexpected nuclear phaseout after the Fukushima Daiichi nuclear disaster. \cite{FUW2012} analyze the stock prices of energy companies using an event study. \cite{BDR2013} use an event study to test for differences in the shareholder wealth of German nuclear energy companies. \cite{T2014} considers the effect on futures prices.

In terms of the research question, our paper is closely related to the recent work of \cite{GHW2017}, who consider differences in the electricity spot prices before and after Germany's nuclear phaseout. Therefore, we use the approach of \cite{GHW2017} as a benchmark for comparing our nonparametric test approach. While \cite{GHW2017} estimate the price differences  conditionally on demand, we additionally allow for an interaction effect with temperature. This way our nonparametric test result demonstrates that a Simpson's paradox \citep{W1982} can explain the unexpected structure of the parametric price differences reported by \cite{GHW2017}.

The literature on covariate-adjusted functional data is fairly scarce. \cite{cardot2007conditional} considers functional principal component analysis for covariate-adjusted random functions, though he focuses on the case of dense functional data and does not derive inferential results for his mean estimate. \cite{jiang2010covariate} show many theoretical results of fundamental importance for sparse covariate-adjusted functional data and we use their pointwise asymptotic normality result for the mean function as a benchmark. \cite{LSB14} consider a copula-based model and \cite{ZW15} propose an iterative algorithm for computing functional principal components, though neither contributes inferential results for the covariate-adjusted mean function. For the case \emph{without} covariate adjustments there are several papers considering inference for the mean function \citep[see, for instance,][]{ZC07,HK07,Benko2009a,CYT12,GK12,HKR13,FS14,VGH14,zhang2016sparse}. We emphasize, however, that the existing results for functional data \emph{without} covariate adjustments cannot easily be generalized to account for additional covariate adjustments. Related to our work is also that of \cite{Serban11} and \cite{GKS2017}, who consider covariate-adjusted, namely, spatio-temporal functional data; however, they do not focus on a sparse functional data context. Two recent works on modeling and forecasting electricity data using methods from functional data analysis are \cite{SL2018} and \cite{LS2018}. Readers with a general interest in functional data analysis are referred to the textbooks of \cite{RamsayfdaBook2005}, \cite{Ferraty2006}, \cite{horvath2012inference}, and \cite{HR15_book}.

The rest of the paper is organized as follows. The next section introduces the LLK estimator and our two-sample test statistic. Section \ref{sec:ROT} introduces  approximations to the unknown bias, variance and bandwidth components. Section \ref{sec:App} contains the real data study. The paper concludes with a discussion in Section \ref{sec:Concl}. The proofs of our theoretical results are based on standard arguments in nonparametric statistics and can be found in the online supplement supporting this article \citep{L17}.

\section{Nonparametric two-sample inference}\label{sec:AssRes}
In the following, we use a common notation for both samples ($A$ and $B$) unless a differentiation is required by the context. Without loss of generality, we consider a standardized domain where $(U_{ij},Z_i)\in[0,1]^2$ such that $X_i(.,Z_i)\in L^2([0,1])$. The standardization can be achieved as $U^{\operatorname{new}}_{ij}=(U^{\operatorname{orig}}_{ij}-a(Z^{\operatorname{orig}}_i))/(b(Z^{\operatorname{orig}}_i)-a(Z^{\operatorname{orig}}_i))$ and $Z^{\operatorname{new}}_i=(Z^{\operatorname{orig}}_i-\min_{1\leq i\leq n}(Z^{\operatorname{orig}}_i))/(\max_{1\leq i\leq n}(Z^{\operatorname{orig}}_i)-\min_{1\leq i\leq n}(Z^{\operatorname{orig}}_i))$. The functional interval borders $a(.)$ and $b(.)$ are unobserved, but can be estimated from the data points $(U_{ij},Z_i)$ using the LLK estimators of \cite{martins2007nonparametric}.

Let $X^c_i$ denote the centered function $X^c_i(u,z)=X_i(u,z)-\E(X_i(u,z))$. Model \eqref{eq:basic_1} can then be rewritten as a nonparametric regression model with the conditional mean function $\mu(U_{ij},Z_i)=\E(X_i(U_{ij},Z_i)|\mathbf{U},\mathbf{Z})$, given $\mathbf{U}=(U_{11},\dots,U_{nm})^\top$ and $\mathbf{Z}=(Z_1,\dots,Z_n)^\top$, i.e.,
\begin{align}\label{Eq2}
  Y_{ij}=\mu(U_{ij},Z_i)+X^c_i(U_{ij},Z_i)+\epsilon_{ij},
\end{align}
where $X^c_i(.,z)$, $U_{ij}$, and $Z_i$ are assumed to be a stationary weakly-dependent functional and univariate time series. The error term $\epsilon_{ij}$ is a classical iid~error term with mean zero, finite variance $\V(\epsilon_{ij})=\sigma^2_\epsilon$, and assumed to be independent from $X^c_s$, $U_{s\ell}$, and $Z_s$ for all $s=1,\dots,n$ and $\ell=1,\dots,m$.

Note that Model \eqref{Eq2} has a rather unusual composed error term consisting of a functional $X^c_i(U_{ij},Z_i)$ and a scalar component $\epsilon_{ij}$. The functional error component introduces very strong local correlations, since 
\begin{align*}
&\operatorname{Corr}\big(X^c_i(U_{ij},Z_i),X^c_i(U_{ik},Z_i)|U_{ij}=u_1,U_{ik}=u_2,Z_i=z\big)=\\
&\operatorname{Corr}\big(X^c_i(u_1,z),X^c_i(u_2,z)\big)\approx 1\quad\text{for}\quad u_1\approx u_2,
\end{align*}
which leads to the above mentioned functional-data-specific variance term that makes it necessary to use the finite sample correction proposed by \citep{DL2018}.

We estimate the mean function $\mu(u,z)$ using the same LLK estimator as considered in \cite{jiang2010covariate}. In the following we define the estimator based on a matrix notion:
\begin{align}
&\hat{\mu}(u,z;h_{\mu,U},h_{\mu,Z})=\label{Estimator_mu}\\
&e_1^{\top}\big([\mathbf{1},\mathbf{U}_u,\mathbf{Z}_z]^{\top}\mathbf{W}_{\mu,uz}[\mathbf{1},\mathbf{U}_u,\mathbf{Z}_z]\big)^{-1}[\mathbf{1},\mathbf{U}_{u},\mathbf{Z}_z]^{\top}\mathbf{W}_{\mu,uz}\mathbf{Y},\notag
\end{align}
where the vector $e_1=(1,0,0)^{\top}$ selects the estimated intercept parameter and $[\mathbf{1},\mathbf{U}_u,\mathbf{Z}_z]$ is a $nm\times 3$ dimensional data matrix with typical rows $(1,U_{ij}-u,Z_i-z)$. The $nm\times nm$ dimensional diagonal weighting matrix $\mathbf{W}_{\mu,uz}$ holds the bivariate multiplicative kernel weights
$K_{\mu,h_{\mu,U},h_{\mu,Z}}(U_{ij}-u,Z_i-z)=h^{-1}_{\mu,U}\,\kappa(h^{-1}_{\mu,U}(U_{ij}-u))\; h^{-1}_{\mu,Z}\,\kappa(h^{-1}_{\mu,Z}(Z_i-z)),$
where $\kappa$ is a usual second-order kernel such as, e.g., the Epanechnikov or the Gaussian kernel, $h_{\mu,U}$ denotes the bandwidth in $U$ direction, and $h_{\mu,Z}$ the bandwidth in $Z$ direction. The kernel constants are denoted by $\nu_{2}(K_\mu)=\left(\nu_{2}(\kappa)\right)^2$, with $\nu_{2}(\kappa)=\int_{[0,1]} u^2\kappa(u)du$, and $R(K_\mu)=R(\kappa)^2$, with $R(\kappa)=\int_{[0,1]}\kappa(u)^2du$. All vectors and matrices are filled in correspondence with the $nm$ dimensional vector $\mathbf{Y}=(Y_{11},Y_{12},\dots,Y_{n,m-1},Y_{nm})^{\top}$.

The following theorem is an adjusted version of Corollary 3.1, part (b) in \cite{DL2018} and takes into account the finite sample correction for the variance component:
\begin{theorem}[Asymptotic normality]\label{C_AN_mu}
Let $m/n^{1/5}\to 0$, let $(u,z)$ be an interior point of $[0,1]^2$, and assume optimal bandwidth rates $h_{\mu,U}\asymp h_{\mu,Z}\asymp (nm)^{-1/6}$. Then the LLK estimator $\hat{\mu}$(u,z) in Eq.~\eqref{Estimator_mu} is asymptotically normal, i.e., 
\begin{align*}
\left(\frac{\hat{\mu}(u,z;h_{\mu,U},h_{\mu,Z})-B_\mu(u,z;h_{\mu,U},h_{\mu,Z})-\mu(u,z)}{\sqrt{V_{\mu}^{\RN{1}}(u,z;h_{\mu,U},h_{\mu,Z})+V_{\mu}^{\RN{2}}(u,z;h_{\mu,Z})}}\right)\overset{a}{\sim}N(0,1),
\end{align*}
where $\mu\in\{\mu_A,\mu_B\}$,\\[-4ex]
\begin{align*}
B_\mu(u,z;h_{\mu,U},h_{\mu,Z})&=\frac{1}{2}\;\nu_{2}(K_\mu)\,\left(h^2_{\mu,U}\,\mu^{(2,0)}(u,z)+h^2_{\mu,Z}\,\mu^{(0,2)}(u,z)\right),\\
V_{\mu}^{\RN{1}}(u,z;h_{\mu,U},h_{\mu,Z})&=\frac{1}{nm}\left[\frac{R(K_\mu)}{h_{\mu,U}h_{\mu,Z}}\frac{\gamma(u,u,z)+\sigma^2_\epsilon}{f_{UZ}(u,z)}\,\right],\\
V_{\mu}^{\RN{2}}(u,z;h_{\mu,Z})&=\frac{1}{n}\left[\left(\frac{m-1}{m}\right)\frac{R(\kappa)}{h_{\mu,Z}}\frac{\gamma(u,u,z)}{f_Z(z)}\right],\quad \text{and}\\
\mu^{(k,l)}(u,z)&=(\partial^{k+l}/(\partial u^k\partial z^l))\mu(u,z).
\end{align*}
This theorem is valid under Assumptions A1-A5, which are listed in the supplement supporting this article \citep{L17}.
\end{theorem}

Theorem \ref{C_AN_mu} generalizes the corresponding result in \cite{DL2018} by additionally allowing for a time series context with weakly dependent auto-correlation structure; a proof can be found in the supplement supporting this article \citep{L17}. The theorem implies the standard optimal convergence rate ($nm^{-1/3}$) for bivariate LLK estimators. The finite sample correction is accomplished by the additional second variance term $V_\mu^{\RN{2}}(u,z;h_{\mu,Z})$ which could be dropped from a pure asymptotic view. However, this second variance term is typically not negligible in practice and serves as a very effective finite sample correction \citep{DL2018}. Note that the variance effects due to the autocorrelations from our time series context are not first-order relevant. The reason for this is that we localize with respect to the variables $U$ and $Z$ and not with respect to time $i$. The resulting decorrelation effect is referred to as the ``whitening window'' property \citep[see, for instance,][Ch.~5.3]{FanTS}.

Theorem \ref{C_AN_mu} without the variance term $V_\mu^{\RN{2}}(u,z;h_{\mu,Z})$, i.e., without finite sample correction, is essentially equivalent to Theorem 3.2 in \cite{jiang2010covariate}, who, however, consider the case where $m$ is bounded, i.e., $1<m\leq c$ for some small $c<\infty$ (e.g., $c=4$ or $c=5$), as typically assumed in the literature on sparse functional data analysis \citep[cf.][and many others]{Yao2005,zhang2016sparse}. In contrast, our asymptotic normality result allows for bounded $m<\infty$ as well as non-bounded $m\to\infty$, such that $m/n^{1/5}\to 0$. This explains the empirical finding in \cite{DL2018} that our asymptotic normality result with finite sample correction provides very good finite sample approximations for practical small-$m$ (e.g., $m=5$) as well as moderate-$m$ cases (e.g., $m=15$)---by contrast to the corresponding results in \cite{jiang2010covariate}.

The following corollary follows directly from Theorem \ref{C_AN_mu} and contains the asymptotic normality result for our two-sample test statistic:
\begin{corollary}[Two-sample test statistic]\label{cor:TSI}
Under the same conditions as in Theorem \ref{C_AN_mu} and under the null hypothesis $H_0$: $\mu_A(u,z)=\mu_B(u,z)$, the following two two-sample test statistic is asymptotically normal:
\begin{align*}
Z_{u,z}=\left(\frac{\hat{\mu}_A(u,z)-B_{\mu_A}(u,z)-\hat{\mu}_B(u,z)+B_{\mu_B}(u,z)}{\sqrt{V^{\RN{1}}_{\mu_A}(u,z)+V^{\RN{2}}_{\mu_A}(u,z)+V^{\RN{1}}_{\mu_B}(u,z)+V^{\RN{2}}_{\mu_B}(u,z)}}\right)\overset{a}{\sim}N(0,1),
\end{align*}
where the dependencies on the bandwidth parameters $h_{\mu,U}$ and $h_{\mu,Z}$ are suppressed for readability reasons.
\end{corollary}

The test statistic $Z_{u,z}$ is infeasible as it depends on the unknown bias, variance, and bandwidth expressions, $B_\mu$, $V_\mu^{\RN{1}}$, $V_\mu^{\RN{2}}$, $h_{\mu,U}$ and $h_{\mu,Z}$. In our application, we use the practical rule-of-thumb bandwidth, bias and variance approximations as described in the following section.

\paragraph*{Remark} A common number $m$ ($m$: number of discretization points per function $X_i$) may be unrealistic for some applications. However, our estimators can be directly applied to data-scenarios with function-specific numbers of discretization points $m_i$ with $i=1,\dots,n$. To adjust our asymptotic analysis for this situation, one can consider the case where $m\to\infty$ with $m\leq m_i$ for all $i=1,\dots,n$ \citep[cf.][]{ZC07}. As \cite{hall2006properties}, \cite{ZC07} and \cite{zhang2016sparse} we do not consider random numbers $m_i$, but if $m_i$ are random, our theory can be considered as conditional on $m_i$.

\section{Practical approximations}\label{sec:ROT}
We approximate the unknown bias term $B_\mu(u,z;h_{\mu,U},h_{\mu,Z})$ by 
\begin{equation}
\begin{array}{l}
\hat{B}_\mu(u,z;h_{\mu,U},h_{\mu,Z})=\\[1ex]
\frac{\nu_2(K_\mu)}{2}
\Big(h_{\mu,U}^2\hat{\mu}^{(2,0)}\big(u,z;g_{\mu,U},g_{\mu,Z}\big)+
h_{\mu,Z}^2\hat{\mu}^{(0,2)}\big(u,z;g_{\mu,U},g_{\mu,Z}\big)\Big),\label{Bias.approx}
\end{array}
\end{equation}
where the estimates of the second-order partial derivatives $\hat{\mu}^{(2,0)}$ and $\hat{\mu}^{(0,2)}$ are local polynomial (order $3$) kernel estimators. That is, 
\begin{equation*}
\begin{array}{l}
\hat{\mu}^{(2,0)}(u,z;g_{\mu,U},g_{\mu,Z})=\\[1ex]
2!\,e_3^{\top}\left([\mathbf{1},\mathbf{U}_u^{1:3},\mathbf{Z}_z^{1:3}]^{\top}\mathbf{W}_{\mu,uz}[\mathbf{1},\mathbf{U}_u^{1:3},\mathbf{Z}_z^{1:3}]\right)^{-1}[\mathbf{1},\mathbf{U}_u^{1:3},\mathbf{Z}_z^{1:3}]^{\top}\mathbf{W}_{\mu,uz}\mathbf{Y}
\end{array}
\end{equation*}
with $e_3^{\top}=(0,0,1,0,0,0,0)$, $\mathbf{U}_u^{1:3}=[\mathbf{U}_u,\mathbf{U}_u^2,\mathbf{U}_u^3]$, $\mathbf{Z}_z^{1:3}=[\mathbf{Z}_z,\mathbf{Z}_z^2,\mathbf{Z}_z^3]$, and diagonal matrix $\mathbf{W}_{\mu,uz}$ with weights $g^{-1}_{\mu,U}\,\kappa(g^{-1}_{\mu,U}(U_{ij}-u))\;g^{-1}_{\mu,Z}\,\kappa(g^{-1}_{\mu,Z}(Z_i-z))$ on its diagonal, where $g_{\mu,U}$ and $g_{\mu,Z}$ are the bandwidths in $U$ and $Z$ direction. The estimator $\hat{\mu}^{(0,2)}$ is defined correspondingly, but with $e_3^\top$ replaced by $e_6^{\top}=(0,0,0,0,0,1,0)$. For estimating the bandwidths $g_{\mu,U}$ and $g_{\mu,Z}$ we use bivariate GCV based on second-order differences. We follow the procedure of \cite{CS15}, but use a GCV-penalty instead of the (asymptotically equivalent) $C_p$-penalty proposed there. 

We estimate the unknown first variance term $V_{\mu}^{\RN{1}}(u,z;h_{\mu,U},h_{\mu,Z})$ by 
\begin{equation}
\begin{array}{l}
\displaystyle
\hat{V}^{\RN{1}}_{\mu}(u,z;h_{\mu,U},h_{\mu,Z},h_{\gamma,U},h_{\gamma,Z})=
\frac{1}{nm}\left[\frac{R(K_\mu)}{h_{\mu,U}h_{\mu,Z}}\frac{\hat{\gamma}^{\text{ND}}(u,u,z;h_{\gamma,U},h_{\gamma,Z})}{\hat{f}_{UZ}(u,z)}\,\right].\label{Var.approx.ND}
\end{array}
\end{equation} 
The Noisy Diagonal (ND) LLK estimator $\hat{\gamma}^{\text{ND}}(u,u,z;h_{\gamma,U},h_{\gamma,Z})$ of $\gamma(u,u,z)+\sigma_\epsilon^2$ is defined as follows:
\begin{equation}
\begin{array}{l}
\displaystyle
\hat{\gamma}^{\text{ND}}(u,u,z;h_{\gamma,U},h_{\gamma,Z})=\\[1ex]
e_1^{\top}\left([\mathbf{1},\mathbf{U}_u,\mathbf{Z}_z]^{\top}\mathbf{W}_{\gamma,uz}[\mathbf{1},\mathbf{U}_u,\mathbf{Z}_z]\right)^{-1}[\mathbf{1},\mathbf{U}_{u},\mathbf{Z}_z]^{\top}\mathbf{W}_{\gamma,uz}\hat{\mathbf{C}}^{\text{ND}},
\end{array}
\end{equation}
with $h_{\gamma,U}$ and $h_{\gamma,Z}$ denoting the bandwidths in $U$ and $Z$ direction and with $\hat{\mathbf{C}}^{\text{ND}}=(\hat C_{111},\dots,\hat C_{ijj}\dots,\hat C_{nmm})^{\top}$ consisting only of the noisy diagonal raw-covariances {\small $\hat{C}_{ijj}^{\text{ND}}=(Y_{ij}-\hat{\mu}(U_{ij},Z_i;h_{\mu,U},h_{\mu,Z}))^2$} for which $\E(\hat C_{ijj}|\mathbf{U},\mathbf{Z})\approx\gamma(U_{ij},U_{ik},Z_i)+\sigma_\epsilon^2$. Note that $\hat{\gamma}^{\text{ND}}$ is equivalent to the LLK estimator ``$\hat{V}$'' in \cite{jiang2010covariate}. The estimate $\hat{f}_{UZ}(u,z)$ is computed using the bivariate kernel density estimation function \texttt{kde2d()} of the \textsf{R}-package \texttt{MASS} \citep{VR2002}, where the involved bandwidths are selected using the \textsf{R}-function \texttt{width.SJ()} of the \textsf{R}-package \texttt{MASS} containing an implementation of the method of \cite{sheather1991reliable}. 

The unknown second variance term $V_{\mu}^{\RN{2}}(u,z;h_{\mu,Z})$ is estimated by
\begin{equation}
\begin{array}{l}
\displaystyle
\hat{V}^{\RN{2}}_{\mu}(u,z;h_{\mu,Z},h_{\gamma,U},h_{\gamma,Z})=\frac{1}{n}\left[\left(\frac{m-1}{m}\right)\frac{R(\kappa)}{h_{\mu,Z}}\frac{\hat{\gamma}(u,u,z;h_{\gamma,U},h_{\gamma,Z})}{\hat{f}_Z(z)}\right].\label{Var.approx.UD}
\end{array}
\end{equation}
The estimate $\hat{f}_{Z}(z)$ is computed using the \textsf{R}-function \texttt{density()} for univariate kernel density estimation, where the involved bandwidth is selected using the \textsf{R}-function \texttt{width.SJ()} of the \textsf{R}-package \text{MASS}. The LLK estimator $\hat{\gamma}(u_1,u_2,z;h_{\gamma,U},h_{\gamma,Z})$ of $\gamma(u_1,u_2,z)$ is defined as follows:
\begin{equation}
\begin{array}{ll}
\displaystyle
&\hat{\gamma}(u_1,u_2,z;h_{\gamma,U},h_{\gamma,Z})=\\
&e_1^{\top}\left([\mathbf{1},\mathbf{U}_{u_{1}},\mathbf{U}_{u_{2}},\mathbf{Z}_z]^{\top}\mathbf{W}_{\gamma,u_{1}u_{2}z}[\mathbf{1},\mathbf{U}_{u_{1}},\mathbf{U}_{u_{2}},\mathbf{Z}_z]\right)^{-1}\times\\
&\hspace{4.1ex}
[\mathbf{1},\mathbf{U}_{u_{1}},\mathbf{U}_{u_{2}},\mathbf{Z}_z]^{\top}\mathbf{W}_{\gamma,u_{1}u_{2}z}\hat{\mathbf{C}}.\label{Estimator_gamma}
\end{array}
\end{equation}
Here, $e_1=(1,0,0,0)^{\top}$ and $[\mathbf{1},\mathbf{U}_{u_{1}},\mathbf{U}_{u_{2}},\mathbf{Z}_z]$ is a $nM\times 4$ dimensional data matrix with typical rows $(1,U_{ij}-u_{1},U_{ik}-u_{2},Z_i-z)$ and $M=m^2-m$. (The latter explains the requirement of Assumption A1 that $m\geq 2$.) The $nM\times nM$ dimensional diagonal weighting matrix $\mathbf{W}_{\gamma,u_{1}u_{2}z}$ holds the trivariate multiplicative kernel weights $K_{\gamma,h_{\gamma,U},h_{\gamma,Z}}(U_{ij}-u_1,U_{ik}-u_2,Z_i-z)=h^{-1}_{\gamma,U}\,\kappa(h^{-1}_{\gamma,U}(U_{ij}-u_1))\; h^{-1}_{\gamma,U}\,\kappa(h^{-1}_{\gamma,U}(U_{ik}-u_2))\;h^{-1}_{\gamma,Z}\,\kappa(h^{-1}_{\gamma,Z}(Z_i-z))$. All vectors and matrices are filled in correspondence with the $nM$ dimensional vector $\hat{\mathbf{C}}=(\hat C_{112},\dots,\hat C_{ijk},\dots,\hat C_{nm,m-1})^{\top}$ consisting only of the off-diagonal raw-covariances
\begin{equation*}
\hat C_{ijk}=(Y_{ij}-\hat\mu(U_{ij},Z_i;h_{\mu,U},h_{\mu,Z}))(Y_{ik}-\hat\mu(U_{ik},Z_i;h_{\mu,U},h_{\mu,Z}))
\end{equation*}
with $j\neq k\in\{1,\dots,m\}$ for which $\E(\hat C_{ijk}|\mathbf{U},\mathbf{Z})\approx\gamma(U_{ij},U_{ik},Z_i)$.
We use bivariate GCV in order to estimate the bandwidth parameters $h_{\mu,U}$, $h_{\mu,Z}$, $h_{\gamma,U}$ and $h_{\gamma,Z}$. 

\section{Application}\label{sec:App}
On March 15, 2011, just after the nuclear meltdown in Fukushima Daiichi, Japan, Germany decided to switch to a renewable energy economy and initiated this by an immediate and permanent shutdown of about $40\%$ of its nuclear power plants. This substantial loss of nuclear power with its low marginal production costs raised concerns about increases in electricity prices and subsequent problems for industry and households. Energy economists typically use Monte Carlo simulations in order to approximate the price effect of Germany's nuclear phaseout \citep[see, for instance,][]{BMDD2013}. Empirical data-based evidence, however, is scarce. \cite{T2014} uses an event study approach to estimate the effect of Germany's nuclear phaseout on electricity futures prices. The work of \cite{GHW2017} considers the less speculative electricity spot prices and can be seen as a parametric counterpart to our work. Therefore, we use the approach of \cite{GHW2017} as a benchmark for our nonparametric approach.

In Section \ref{ssec:BR} we introduce the benchmark models. In Section \ref{ssec:NTR} we compare the benchmark results with the results of our nonparametric two-sample test statistic and demonstrate that a Simpson's paradox can explain the unexpected finding in \cite{GHW2017}.

\paragraph*{Data} The data for our analysis come from different sources that are described in detail in the online supplement supporting this article \citep{L17}. The German electricity market, like many others, provides purchase guarantees for renewable energy sources. Therefore, the relevant variable for pricing is electricity demand (or ``load'') minus electricity infeeds from RES and an additional correction for the net imports of electricity from neighboring countries \citep[see, e.g.,][]{PEP2014}. Correspondingly, in our application 
$U_{ij}$ refers to \emph{residual} electricity demand defined as $U_{ij}=D_{ij}-R_{ij}+N_{ij}$, with $R_{ij}=W_{ij}+S_{ij}$ and $N_{ij}=I_{ij}-E_{ij}$, where $D_{ij}$ denotes electricity demand, $R_{ij}$ denotes infeeds from renewable energy sources, $W_{ij}$ denotes wind-power, $S_{ij}$ denotes solar-power, $N_{ij}$ denotes net-imports, $I_{ij}$ denotes electricity imports, and $E_{ij}$ denotes electricity exports.  
The effect of further renewable energy sources such as biomass is still negligible for the German electricity market. Very few ($0.2\%$) of the data tuples $(Y_{ij},U_{ij},Z_i)$ with prices $Y_{ij}>200$ EUR/MWh are considered as outliers and set to $Y_{ij}=200$ EUR/MWh. Such extreme prices are often referred to as ``price spikes''; they are caused by market speculations involving potential capacity scarcities and need to be modeled using different approaches \citep[see, for instance,][Ch.~4]{burger2004spot}. Our data set consists of the peak-hour prices ($m=12$; from $9$am to $8$pm) of the working days from one year before ($n_B=242$) and one year after ($n_A=239$) Germany's partial nuclear phaseout on March 15, 2011. 
We consider only working days, since for weekends there are different compositions of the power plant portfolio. The same reasoning applies to holidays and so-called Br\"uckentage, which are extra days off that bridge single working days between a bank holiday and the weekend. Therefore, we remove also all holidays and Br\"uckentage from the data. 
The temporal gaps in the data due to weekend days, holidays and Br\"uckentage do not violate our theoretical assumptions on the auto-covariance structure, since such gaps do not increase the auto-correlations. 
The main fundamental market factors (i.e., temperature, gas, CO$_2$ allowance, and coal prices) are available at a daily sampling scheme. Legal issues do not allow us to publish the original data sets, however, simulated data sets that closely resemble the original data can be found in the online supplement supporting this article \citep{L17}.

\subsection[]{Parametric benchmark models}\label{ssec:BR}
As a benchmark case study for our nonparametric approach, we use the following two increasingly complex parametric regression models:
{\small\begin{align}
Y_i&=\alpha_1 + \alpha_2 d_i + \alpha_3 Z_i    + \sum_{k=1}^K\beta_k \mathfrak{X}_{ik} + \epsilon_i,\label{eq:BM1}\\
Y_i&=\alpha_1 + \alpha_2 U_i + \alpha_3 U_i^2  + d_i\big(\alpha_4 + \alpha_5 U_i + \alpha_6 U_i^2\big) + \alpha_7 Z_i +\
\sum_{k=1}^K\beta_k \mathfrak{X}_{ik} + \epsilon_i,\label{eq:BM2}
\end{align}} 
\vspace*{-3mm}
 
\noindent where $Y_i=12^{-1}\sum_{j=1}^{12}Y_{ij}$ denotes the daily mean (peak-hours) electricity spot price, $U_i=12^{-1}\sum_{j=1}^{12}U_{ij}$ denotes the daily mean (peak-hours) Residual Demand (RD), $d_{i}$ is a dummy variable which equals zero for all time points before Germany's nuclear phaseout and equals one for all other time points, $Z_i$ denotes air temperature, $\mathfrak{X}_{ik}$ contains further control variables, and $\epsilon_i$ is a classical Gaussian statistical error term. As control variables $\mathfrak{X}_{ik}$ we use the same fundamental market factors as in the fundamental electricity market model of \cite{MW2016}, i.e., temperature, CO$_2$ emission allowance prices, coal prices, and natural gas prices. For lignite and nuclear energy resources there are no relevant market prices \citep{GHW2017}. 

\begin{figure}[!ht]
\centering
\includegraphics[width=1\textwidth]{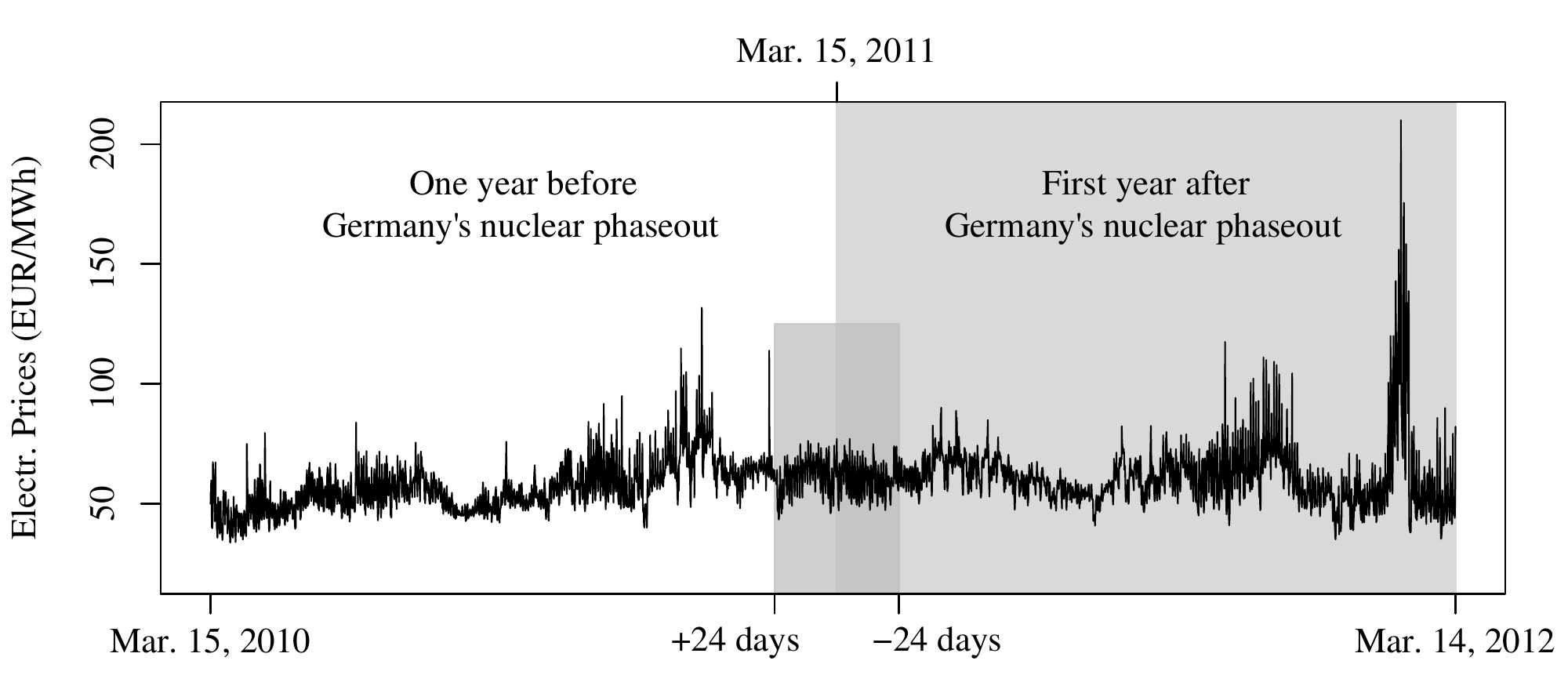}
\caption[]{Time series of Germany's hourly electricity spot prices.}
\label{Fig:TSP} 
\end{figure} 
Model \eqref{eq:BM1} estimates a possible price effect using a simple dummy variable approach. Model \eqref{eq:BM2} corresponds to the regression model of \cite{GHW2017}, who estimate the price effect of Germany's nuclear phaseout using the quadratic dummy-interaction function $d_i(\alpha_4 + \alpha_5 x_i + \alpha_6 x_i^2)$; see Table 4 in \citealp{GHW2017}. We cannot consider most all of the control variables proposed by \cite{GHW2017}, since we do not have access to the data used in their case study (missing variables: export congestion index, residual supply index (i.e., a market power index), and low/high river level). In order to rectify this shortcoming, we use the event study approach with short estimation and event windows each containing 24 days (see Figure \ref{Fig:TSP}). Our window size is one day smaller than the window size in the related event study of \cite{T2014}, since we want to exclude the strong price jump one day before our estimation window. Our estimation results are robust to smaller window sizes. 

\begin{table}[!htbp] 
\centering 
\caption{Event study estimation results for the parametric benchmark Models \eqref{eq:BM1} and \eqref{eq:BM2}.}\label{tab:TSBM} 
\begin{tabular}{@{\extracolsep{0pt}}l  D{.}{.}{-3} D{.}{.}{-3} D{.}{.}{-3} D{.}{.}{-3} c D{.}{.}{-3} D{.}{.}{-3} D{.}{.}{-3} D{.}{.}{-3} } 
\toprule \\[-1.8ex] 
& \multicolumn{4}{c}{Model \eqref{eq:BM1}} && \multicolumn{4}{c}{Model \eqref{eq:BM2}} \\ 
\cmidrule{2-5}\cmidrule{7-10}\\[-2ex] 
&\multicolumn{1}{c}{I}&\multicolumn{1}{c}{II}&\multicolumn{1}{c}{III}&\multicolumn{1}{c}{IV}&
&\multicolumn{1}{c}{I}&\multicolumn{1}{c}{II}&\multicolumn{1}{c}{III}&\multicolumn{1}{c}{IV}\\
RD                                           &       &       &        &        && 0.0   & 0.0   & 0.0    & 0.0    \\ 
RD$^2$                                       &       &       &        &        &&-0.0   &-0.0   &-0.0    &-0.0    \\ 
\rowcolor{black!20}Dummy                     &-1.1   & 1.9   & 5.7^{*}& 2.6^{*}&&63.3   &88.2   &61.3    &42.1    \\ 
\rowcolor{black!20}Dummy$\times$RD           &       &       &        &        &&-0.0   &-0.0   &-0.0    &-0.0    \\ 
\rowcolor{black!20}Dummy$\times$RD$^2$       &       &       &        &        && 0.0   & 0.0   & 0.0    & 0.0    \\ 
Temperature                                  &       &       &-1.3^{*}&-1.2^{*}&&       &       &-0.8^{*}&-0.8^{*}\\ 
Temperature$^2$                              &       &       & 0.1^{*}& 0.0    &&       &       & 0.0    & 0.0^{*}\\ 
CO$_2$\hspace*{.2mm} price                   &       & 0.4   &-1.4    &        &&       & 2.2   & 0.7    &        \\ 
Coal price                                   &       &-0.7   &-0.2    &        &&       & 0.2   & 0.4    &        \\ 
Gas\phantom{l} price                         &       & 0.8   & 0.2    &        &&       &-0.9   &-1.3    &        \\ 
\midrule  
R$^{2}$                                      & 0.0   & 0.1   & 0.4    & 0.3    && 0.5   & 0.6   & 0.6    & 0.6    \\ 
Adj.~R$^{2}$                                 & 0.0   & 0.0   & 0.3    & 0.3    && 0.5   & 0.5   & 0.5    & 0.5    \\ 
\midrule  
F Statistics                                 & 1.1   & 1.5   & 4.2^{*}& 7.6^{*}&&9.0^{*}&6.0^{*}& 6.4^{*}& 8.6^{*}\\ 
I\phantom{II} vs II                          &  \multicolumn{2}{c}{1.6}     && &&   \multicolumn{2}{c}{1.0}     &&\\ 
II\phantom{I} vs III                         && \multicolumn{2}{c}{8.8$^{*}$}& && & \multicolumn{2}{c}{4.1$^{*}$}&\\  
III           vs IV                          &&&\multicolumn{2}{c}{0.9}        && &&\multicolumn{2}{c}{1.1}       \\  
\bottomrule \\[-1.8ex] 
\multicolumn{10}{l}{\textit{Note:} $^{*}$p$<$0.05} 
\end{tabular} 
\end{table} 

The F statistics in the lower panel of Table \ref{tab:TSBM} show that the simple benchmark Model \eqref{eq:BM1} is insufficient for describing the price effect of Germany's nuclear phaseout, expect if the temperature-component is added to the control variables (III and IV). In contrast to this, Model \eqref{eq:BM2}, which contains the quadratic dummy-interaction function, describes the price effect with a significantly strong explanatory power in all model-specifications I-IV. Comparing the different model-specifications reveals that the resource prices (CO$_2$, coal, and gas) are jointly insignificant control variables: neither adding them to the smallest model-specification (I vs II), nor removing them from the larges model-specification (III vs IV) has a significant impact on the fit of the models. That is, our event study approach successfully minimizes the confounding effects of the control variables CO$_2$, coal, and gas, which are known to be less important in the short run. Furthermore, it demonstrates the importance of including the temperature-component.

The graph of the estimated quadratic dummy-interaction function of benchmark Model (\ref{eq:BM2}-IV) together with its 95\% confidence interval is shown in Figure \ref{Fig:ETR}. The confidence interval covers the dummy-interaction function reported\footnote{The data for this graph are extracted from Figure 6 in \cite{GHW2017} using the WebPlotDigitizer of \cite{R2018}.} by \cite{GHW2017}. The dummy-interaction function of our benchmark Model shows slightly larger values than reported by \cite{GHW2017}, since the market reactions to Germany's unexpected nuclear phaseout are less moderated in the short time frame of our event study than in the long time frame ($\pm 2$ years) considered by \cite{GHW2017}.

\subsection{Nonparametric testing}\label{ssec:NTR}
In contrast to the parametric benchmark models, we do not assume any specific functional form for the price effect of Germany's nuclear phaseout. Additionally, we allow for interactions between residual demand and temperature. As for any nonparametric model, however, we need to deal with the curse of dimensionality which prevents us from considering all control variables of potential relevance. Therefore, we focus on the two most important control variables (residual demand and temperature) and use the event study approach in order to minimize the confounding effects of the remaining less important control variables as in our benchmark study.  


Residual demand and temperature are the two most important market fundamental control variables within the short term of an event study. First, because the interplay of residual demand with the merit-order curve determines electricity spot prices (see Figure \ref{Fig:MOC}). Second, because the merit-order curve itself depends on temperature. For the latter dependency, there are multiple reasons, both fundamental and speculative. The most important fundamental reason is that the merit-order curve is determined by the generation costs of the conventional (i.e., nuclear, lignite, coal, and gas) power plants. These conventional power plants are thermal-based electricity generators using river water as their primary cooling resource. The resulting thermal pollution is substantial and environmentally hazardous and, therefore, strictly regulated by public institutions. This regulation affects the cost structure, i.e., the merit-order curve \citep{McD2014}. For given values of electricity demand, the merit-order curve is higher (lower) on warm (cold) days, when the river water has a smaller (larger) cooling capacity\footnote{\cite{McD2014} propose to use river temperature as a control variable. However, we do not have access to their data and therefore use air temperature as a proxy which is known to correlate strongly with river temperature \citep[see][]{RHS2015}} (see Figure \ref{Fig:SP}, for middle to high temperatures). However, there is also a reverse interaction effect between residual demand and temperature which can revert the fundamental price component described above. Due to the use of heating and cooling devices, cold and hot days are associated with the large amounts of electricity demand resulting in market situations where the electricity production capacities become scarce.\footnote{The use of air conditioning systems is, however, less extensive in Germany than, for instance, in the US.} In these situations of market stress, one observes the high electricity prices (see Figure \ref{Fig:SP}, for cold temperatures), since the electricity producers demand additional scarcity premiums \citep[see][Ch.~4]{Burger2008}.

To conclude, the effects of residual demand and temperature and their interaction effects are quite complex and it is easy to end up with a misspecified model, particularly when using a parametric approach. Despite these complexities, the parametric benchmark Model \eqref{eq:BM2}, proposed by \cite{GHW2017}, does not consider interaction effects between electricity demand and temperature. In the following, we compare our nonparametric test results with our parametric benchmarks and demonstrate the importance of this interaction effect. 

\begin{figure}[!ht]   
\centering   
\includegraphics[width=1\textwidth]{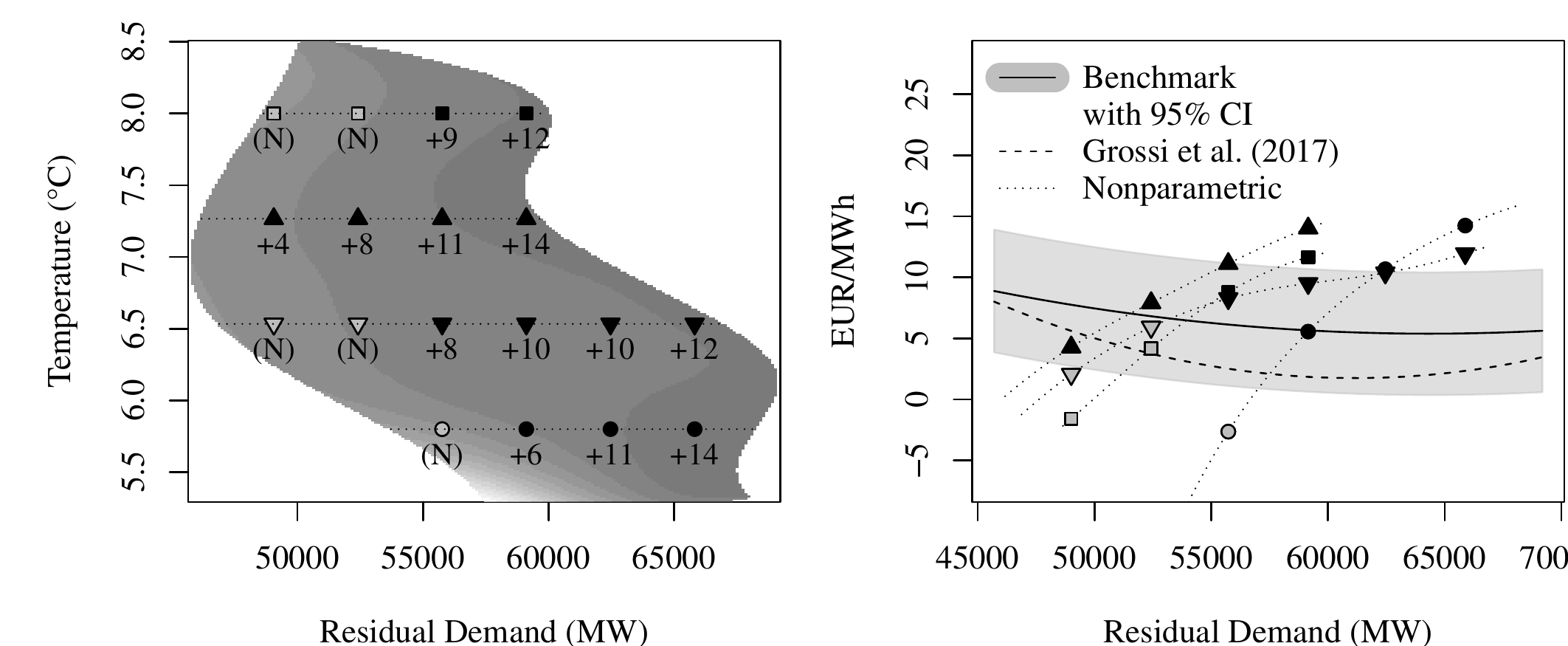}
\caption[]{{\sc Left:} Contour plot of the price difference surface $\hat\mu_A(u,z)-\hat\mu_B(u,z)$; (N) denotes insignificant results. {\sc Right:} Comparisons of the results from benchmark Model (\ref{eq:BM2}-IV), the original result of \cite{GHW2017}, and our nonparametric result.}   
\label{Fig:ETR}
\end{figure}  

The left plot in Figure \ref{Fig:ETR} shows a contour plot of the price difference surface $\hat{\mu}_{B}-\hat{\mu}_{A}$ of the mean estimates before and after Germany's nuclear phaseout. The support of the difference surface equals the intersection $\operatorname{supp}(\hat{\mu}_{B})\cap\operatorname{supp}(\hat{\mu}_{A})$ of the supports of the mean functions $\hat{\mu}_{B}$ and $\hat{\mu}_{A}$, 
\begin{align*}
\operatorname{supp}(\hat{\mu}_{B})=&\{(u,z):\hat a_{B}(z)\leq u\leq \hat b_{B}(z)\;\text{ with }\;\hat z_{\min,B}\leq z\leq \hat z_{\max,B}\}\quad\text{and}\\
\operatorname{supp}(\hat{\mu}_{A})=&\{(u,z):\hat a_{A}(z)\leq u\leq \hat b_{A}(z)\;\text{ with }\;\hat z_{\min,A}\leq z\leq \hat z_{\max,A}\},
\end{align*}
where the empirical boundary functions $\hat a_P(.)$ and $\hat b_P(.)$, $P\in\{A,B\}$, are computed using the LLK boundary estimator of \cite{martins2007nonparametric}.

In order to test the pointwise null hypothesis $\operatorname{H_0}$: $\mu_A(u,z)=\mu_B(u,z)$ against the alternative $\operatorname{H_1}$: $\mu_A(u,z)>\mu_B(u,z)$, we use our finite sample corrected two-sample test statistic $Z_{u,z}$ described in Section \ref{sec:AssRes} with plugged-in empirical bias and variance expressions and GCV-optimal bandwidths as described in Section \ref{sec:ROT}. The test statistic is evaluated at $G=18$ regular grid-points $(u_j,z_j)$, $j=1,\dots,G$, within the intersection $\operatorname{supp}(\hat{\mu}_{B})\cap \operatorname{supp}(\hat{\mu}_{A})$. These test-points are shown by the points in the plots of Figure \ref{Fig:ETR}, where the different point shapes correspond to different temperature values. In order to account for the multiple testing we use a Bonferroni-adjusted significance level $\alpha/G$ where we set $\alpha=0.05$. Significant differences at the chosen test-points are depicted by the numerical values in the left plot of Figure \ref{Fig:ETR}; non-significant differences are marked by ``(N)''. We are interested in pointwise hypotheses as we would like to identify significant and non-significant test-points. The Bonferroni-adjustment is known to be conservative, but works well for our application. Practitioners seeking for an alternative to the Bonferroni-adjustment are referred, for instance, to the work of \cite{CL2008}.

The plots in Figure \ref{Fig:ETR} show that the price differences are large (moderate) for large (moderate) values of electricity demand. This is in line with our expectations, since the merit-order curve is known to be steep (relatively flat) for large (moderate) values of electricity demand resulting in large (moderate) price differences (see Figure \ref{Fig:MOC}). Furthermore, there is a clear interaction effect in the price differences. If temperature increases, the thermal power plans loose cooling capacities which increases the production costs---particularly for the less efficient power plants. The latter results in a stepper merit-order curve and, therefore, larger price differences.

\section{Discussion}\label{sec:Concl}
On March 15, 2011, the German government showed a drastic reaction to the nuclear disaster in Fukushima Daiichi, Japan, and permanently shut down $40\%$ of its nuclear power plants. This political decision raised concerns about increases in electricity prices and subsequent problems for industry and households. Empirical studies on possible price effects, however, are scarce and existing studies are based on restrictive parametric model assumptions.

In this work we add a functional data perspective based on the merit-order model, the most important model for explaining electricity spot prices \citep[see][]{burger2004spot, sensfuss2008merit, hirth2013market, L13, cludius2014merit, BKF17, GHW2017}. We extend the work of \cite{L13} and additionally control for non-functional covariate adjustments.

In order to test for a possible price effect, we compare the multivariate nonparametric local linear kernel estimates of the mean price functions before and after Germany's nuclear phase out on March 15, 2011, using a pointwise test statistic. Nonparametric smoothing of the pooled data is used, since the underlying daily price functions are only observed at $12$ noisy discretization points (``sparse functional data''). The existing asymptotic results on this nonstandard smoothing problem only consider the leading variance term and neglect an additional functional data specific variance term, which is asymptotically negligible, but typically not practically negligible. Ignoring this additional variance term can result in serious size distortions and invalid test decisions \citep[see][]{DL2018}. Therefore, we propose a finite sample correction that considers also the second functional data  specific variance term (see Theorem \ref{C_AN_mu} and Corollary \ref{cor:TSI}). Theorem \ref{C_AN_mu} generalizes the main result in \cite{DL2018} by allowing for a time series context with weak dependency structure.

We compare our nonparametric test results with parametric benchmark results replicating the results recently reported by \cite{GHW2017}. Our results confirm the existence of a price effect due to Germany's abrupt nuclear phase out, but our price effect is structured quite differently to the results in \cite{GHW2017}. While our nonparametric price differences are highest for large values of residual demand, the parametric benchmark models estimate the highest price differences for small values of residual demand (see right panel in Figure \ref{Fig:ETR}). One of the fundamental differences between our approach and the approach of \cite{GHW2017} is that we take into account interactions with the important temperature factor, while \cite{GHW2017} do not allow for this kind of interaction effect. Such a reversal of effects that is due to introducing an additional conditioning variable is known as Simpson's paradox \citep[see, for instance,][]{W1982}. \cite{GHW2017} concede that their result is ``unexpected'' and present different market explanations for their unexpected result; however, a possible model-misspecification is not taken into account. Our nonparametric case study points at such a possible model-misspecification and demonstrates that a Simpson's paradox can explain their unexpected finding.

\section*{Acknowledgements}
I would like to thank my colleague Alois Kneip (University of Bonn) for many stimulating and fruitful discussions on this project. Furthermore, I am grateful to the referees and the editors for their constructive feedback which helped to improve this research work. 

\bigskip

\begin{supplement}
\sname{Supplement A}
\stitle{\textsf{R}-Codes and Data}
\slink[doi]{COMPLETED BY THE TYPESETTER}
\sdatatype{zip file}
\sdescription{This supplementary material contains the \textsf{R} codes of the real data application and simulated data which closely resembles the original data set.}
\end{supplement}
\begin{supplement}
\sname{Supplement B}
\stitle{Supplementary Paper}
\slink[doi]{COMPLETED BY THE TYPESETTER}
\sdatatype{pdf file}
\sdescription{This supplementary paper contains the proofs of our theoretical results and a detailed description of the data sources.}
\end{supplement}

\newpage

\bibliographystyle{imsart-nameyear}
\bibliography{bibfile.bib}

\hfilneg\printaddressnum{1}

\newpage
\setcounter{page}{1}
\pagenumbering{Roman}
\setcounter{section}{0}
\thispagestyle{empty}
\vspace*{.5cm}
\noindent{\large \bf Supplementary Paper for:}\\
\begin{center}
{\Large \bf Nonparametric Testing for Differences in Electricity Prices: The Case of the Fukushima Nuclear Accident}\\[4ex]
{\sc By Dominik Liebl}\\
\end{center}
\appendix

Our basic assumptions, are essentially equivalent to those in \cite{ruppert1994} with some straightforward adjustments to our functional data and time series context.
\begin{description}
\item{A1} (Asymptotic Scenario) $nm\to\infty$, where $m=m_n\geq 2$ such that $m_n\asymp n^\theta$ with $0\leq\theta<\infty$. Hereby, ``$m_n\asymp n^\theta$'' denotes that the two sequences $m_n$ and $n^\theta$ are asymptotically equivalent, i.e., that $\lim_{n\rightarrow\infty}(m_n/ n^\theta)=C$ with constant $0<C<\infty$.

\item{A2} (Random Design)
The triple $(Y_{ij}, U_{ij}, Z_i$) has the same distribution as $(Y,U,Z)$ with pdf~$f_{YUZ}$ where $f_{YUZ}(y,u,z)>0$ for all $(y,u,z)\in\mathbb{R}\times[0,1]^2$ and zero else.
The error term $\epsilon_{ij}$ is iid and independent from $X^c_s$, $U_{s\ell}$, and $Z_s$ for all $s=1,\dots,n$ and $\ell=1,\dots,m$. 

\item{A3} (Smoothness \& Kernel)
The pdf $f_{YUZ}(y,u,z)$ and its marginals are continuously differentiable. All second-order derivatives of the function $\mu$ are continuous. 
The (auto-)covariance functions $\gamma_{l}((u_1,z_1),(u_2,z_2))=\E(X^c_i(u_1,z_1)X^c_{i+l}(u_2,z_2))$, $l\geq 0$, are continuously differentiable for all points within their supports.
The multiplicative kernel functions $K_\mu$ and $K_\gamma$ are products of second-order kernel functions $\kappa$.

\item{A4} (Moments \& Dependency)
$X_i$, $U_{ij}$, and $Z_i$ are strictly stationary, ergodic, and weakly dependent time series with auto-covariances that converge uniformly to zero at a geometrical rate. It is assumed that $\E(X_i(u,z)^4)<\infty$, $\E(\epsilon_{ij})=0$, $\E(\epsilon_{ij}^2)=\sigma^2_\epsilon<\infty$ for all $(u,z)$, $i$, and $j$.

\item{A5} (Bandwidths)
$h_{\mu,U},h_{\mu,Z}\to 0$ and
$(nm)h_{\mu,U}h_{\mu,Z}\to\infty$ as $nm\to\infty$.
$h_{\mu,U},h_{\mu,Z}\to 0$ and
$(nM)h_{\mu,U}^2h_{\mu,Z}\to\infty$ as $nM\to\infty$.

\end{description}

\smallskip

\paragraph{Remark} Assumption A1 is a simplified version of the asymptotic setup of \cite{zhang2016sparse}. The case $\theta=0$ implies that $m$ is bounded which corresponds to a simplified version of the finite-$m$ asymptotic considered by \cite{jiang2010covariate}. For $0<\theta<\infty$ we can consider all further scenarios from sparse to dense functional data. In line with our real data application, we consider a deterministic $m$ as also done, for instance, by \cite{hall2006properties}. However, our results are generalizable to a random $m$ using some minor modifications.

\section{Proofs}\label{ProofTheoremBiasVar}
The following Lemma \ref{Bias_and_Variance_mu} builds the basis of our theoretical results.
\begin{lemma}[Bias and Variance of $\hat{\mu}$]\label{Bias_and_Variance_mu}
Let $(u,z)$ be an interior point of $[0,1]^2$. Under Assumptions A1-A5 the conditional asymptotic bias and variance of the LLK estimator $\hat{\mu}$ in Eq.~\eqref{Estimator_mu} are then given by
\begin{align*}
&\text{\textit{(i)}} \Bias\left\{\hat{\mu}(u,z;h_{\mu,U},h_{\mu,Z})|\mathbf{U},\mathbf{Z}\right\}=B_\mu(u,z)+o_p(h^2_{\mu,U}+h^2_{\mu,Z})\text{ with}\\
&B_\mu(u,z)=\frac{1}{2}\;\nu_{2}(K_\mu)\,\left(h^2_{\mu,U}\,\mu^{(2,0)}(u,z)+h^2_{\mu,Z}\,\mu^{(0,2)}(u,z)\right),\text{ where}\\
&\mu^{(k,l)}(u,z)=(\partial^{k+l}/(\partial u^k\partial
z^l))\mu(u,z).\\[2ex]
&\text{\textit{(ii)} }
\V\left\{\hat{\mu}(u,z;h_{\mu,U},h_{\mu,Z})|\mathbf{U},\mathbf{Z}\right\}=\left(V_\mu^{\RN{1}}(u,z)+V_\mu^{\RN{2}}(u,z)\right)\left(1+o_p(1)\right)\text{with}\\
&V_\mu^{\RN{1}}(u,z)=(nm)^{-1}\left[h_{\mu,U}^{-1}h_{\mu,Z}^{-1}\,R(K_\mu)\frac{\gamma(u,u,z)+\sigma^2_\epsilon}{f_{UZ}(u,z)}\,\right]\text{and}\\
&V_\mu^{\RN{2}}(u,z)=n^{-1}\left[\left(\frac{m-1}{m}\right)h_{\mu,Z}^{-1}\,R(\kappa)\frac{\gamma(u,u,z)}{f_Z(z)}\right].
\end{align*}
\end{lemma}

\smallskip\noindent\textbf{Proof of Lemma \ref{Bias_and_Variance_mu}.}
Our proof of Lemma \ref{Bias_and_Variance_mu} generally follows that of \citeappendix{RW94}, and differs only from the latter reference as we consider additionally a conditioning variable $Z_i$, a function-valued error term, and a time series context.

Proof of Lemma \ref{Bias_and_Variance_mu}, part \textit{(i)}: For simplicity, consider a second-order kernel function $\kappa$ with compact support such as the Epanechnikov kernel; this is, of course, without loss of generality. Let $(u,z)$ be a interior point of $[0,1]^2$ and define $\mathbf{H}_\mu=\mydiag(h^2_{\mu,U},h^2_{\mu,Z})$, $\mathbf{U}=(U_{11},\dots,U_{nm})^\top$, and $\mathbf{Z}=(Z_1,\dots,Z_n)^\top$. Using a Taylor-expansion of $\mu$ around $(u,z)$, the conditional bias of the estimator $\hat{\mu}(u,z;\mathbf{H})$ can be written as
\begin{align}
&\E(\hat{\mu}(u,z;\mathbf{H}_\mu)-\mu(u,z)|\mathbf{U},\mathbf{Z})=\label{Bias1}\\
&=\frac{1}{2}e_1^{\top}\left((nm)^{-1}[\mathbf{1},\mathbf{U}_{u},\mathbf{Z}_{z}]^{\top}\mathbf{W}_{\mu,uz}[\mathbf{1},\mathbf{U}_{u},\mathbf{Z}_{z}]\right)^{-1}\times\notag\\
&\times(nm)^{-1}[\mathbf{1},\mathbf{U}_{u},\mathbf{Z}_{z}]^{\top}\mathbf{W}_{\mu,uz}\left(\boldsymbol{\mathcal{Q}}_\mu(u,z)+\mathbf{R}_\mu(u,z)\right),\notag
\end{align}
where $\boldsymbol{\mathcal{Q}}_\mu(u,z)$ is a $nm\times 1$ vector with typical elements
\begin{eqnarray*}
  (U_{ij}-u,Z_i-z)\boldsymbol{\mathcal{H}}_\mu(u,z)(U_{ij}-u,Z_i-z)^{\top}\in\mathbb{R}
\end{eqnarray*}
with $\boldsymbol{\mathcal{H}}_\mu(u,z)$ being the Hessian matrix of the regression function $\mu(u,z)$. The $nm\times 1$ vector $\mathbf{R}_\mu(u,z)$ holds the remainder terms as in \citeappendix{RW94}.

Next we derive asymptotic approximations for the $3\times 3$ matrix\\
$\left((nm)^{-1}[\mathbf{1},\mathbf{U}_{u},\mathbf{Z}_{z}]^{\top} \mathbf{W}_{\mu,uz}[\mathbf{1},\mathbf{U}_{u},\mathbf{Z}_{z}]\right)^{-1}$ and the $3\times 1$ matrix \\
$(nm)^{-1}[\mathbf{1},\mathbf{U}_{u},\mathbf{Z}_{z}]^{\top}\mathbf{W}_{\mu,uz}\boldsymbol{\mathcal{Q}}_\mu(u,z)$ of the right hand side of Eq.~\eqref{Bias1}. Using standard arguments from nonparametric statistics it is easy to derive that \\
$(nm)^{-1}[\mathbf{1},\mathbf{U}_{u},\mathbf{Z}_{z}]^{\top}\mathbf{W}_{\mu,uz}[\mathbf{1},\mathbf{U}_{u},\mathbf{Z}_{z}]=$
{\small\begin{eqnarray*}
  \left(\begin{matrix}
      f_{UZ}(u,z)+o_p(1)                                      &\quad \nu_{2}(K_\mu)\mathbf{D}_{f_{UZ}}(u,z)^{\top}\mathbf{H}_\mu+o_p(\mathbf{1}^\top\mathbf{H}_\mu)\\
      \nu_{2}(K_\mu)\mathbf{H}_\mu \mathbf{D}_{f_{UZ}}(u,z)+o_p(\mathbf{H}_\mu\mathbf{1})&\quad \nu_{2}(K_\mu)\mathbf{H}_\mu f_{UZ}(u,z)+o_p(\mathbf{H}_\mu)\\
    \end{matrix}\right),
\end{eqnarray*}}\noindent 
where $\mathbf{1}=(1,1)^\top$ and $\mathbf{D}_{f_{UZ}}(u,z)$ is the vector of first order partial derivatives (i.e., the gradient) of the pdf $f_{UZ}$ at $(u,z)$. Inversion of the above block matrix yields
\begin{eqnarray}\label{pre_results}
  \left((nm)^{-1}[\mathbf{1},\mathbf{U}_{u},\mathbf{Z}_{z}]^{\top}\mathbf{W}_{\mu,uz}[\mathbf{1},\mathbf{U}_{u},\mathbf{Z}_{z}]\right)^{-1}=&&
\end{eqnarray}
{\small\begin{eqnarray*}
  \left(\begin{matrix}
      \left(f_{UZ}(u,z)\right)^{-1}+o_p(1)                          &\quad  -\mathbf{D}_{f_{UZ}}(u,z)^{\top}\left(f_{UZ}(u,z)\right)^{-2}+o_p(\mathbf{1}^\top)\\
      -\mathbf{D}_{f_{UZ}}(u,z)\left(f_{UZ}(u,z)\right)^{-2}+o_p(\mathbf{1}) &\quad \left(\nu_{2}(K_\mu)\mathbf{H}_\mu f_{UZ}(u,z)\right)^{-1}+o_p(\mathbf{H}_\mu)\\
\end{matrix}\right).
\end{eqnarray*}}
The $3\times 1$ matrix $(nm)^{-1}[\mathbf{1},\mathbf{U}_{u},\mathbf{Z}_{z}]^{\top}\mathbf{W}_{\mu,uz}\boldsymbol{\mathcal{Q}}_\mu(u,z)$ can be partitioned as following:
\begin{eqnarray*}
  (nm)^{-1}[\mathbf{1},\mathbf{U}_{u},\mathbf{Z}_{z}]^{\top}\mathbf{W}_{\mu,uz}\boldsymbol{\mathcal{Q}}_\mu(u,z)&=&\left(\begin{matrix}\texttt{upper element}\\
      \texttt{lower bloc}
    \end{matrix}\right),
\end{eqnarray*}
where the $1\times 1$ dimensional \texttt{upper element} can be approximated by
{\small\begin{align}
  &(nm)^{-1}\sum_{ij}K_{\mu,h}(U_{ij}-u,Z_{\RN{1}}-z)(U_{ij}-u,Z_{\RN{1}}-z)\boldsymbol{\mathcal{H}}_\mu(u,z)(U_{ij}-u,Z_{\RN{1}}-z)^{\top}\label{Q1}\\
  =&\left(\nu_{2}(\kappa)\right)^2tr\left\{\mathbf{H}_\mu\boldsymbol{\mathcal{H}}_\mu(u,z)\right\}f_{UZ}(u,z)+o_p(tr(\mathbf{H}_{\mu}))\notag
\end{align}}
and the $2\times 1$ dimensional \texttt{lower bloc} is equal to
{\small\begin{align}
&(nm)^{-1}\sum_{ij}\left\{K_{\mu,h}(U_{ij}-u,Z_{\RN{1}}-z)(U_{ij}-u,Z_{\RN{1}}-z)\boldsymbol{\mathcal{H}}_\mu(u,z)(U_{ij}-u,Z_i-z)^{\top}\right\}\times\label{Q2}\\
&\times(U_{ij}-u,Z_{\RN{1}}-z)^{\top}=O_p(\mathbf{H}_\mu^{3/2}\mathbf{1}).\notag
\end{align}}\noindent
Plugging the approximations of Eqs.~\eqref{pre_results}-\eqref{Q2} into the first summand of the conditional bias expression in Eq.~\eqref{Bias1} leads to the following expression
\begin{align*}
&\frac{1}{2}e_1^{\top}\big((nm)^{-1}[\mathbf{1},\mathbf{U}_{u},\mathbf{Z}_{z}]^{\top}\mathbf{W}_{\mu,uz}[\mathbf{1},\mathbf{U}_{u},\mathbf{Z}_{z}]\big)^{-1}\times\\
&\times(nm)^{-1}[\mathbf{1},\mathbf{U}_{u},\mathbf{Z}_{z}]^{\top}\mathbf{W}_{\mu,uz}\boldsymbol{\mathcal{Q}}_\mu(u,z)=\\
&=\frac{1}{2}\left(\nu_{2}(\kappa)\right)^2tr\left\{\mathbf{H}_\mu\boldsymbol{\mathcal{H}}_\mu(u,z)\right\}+o_p(tr(\mathbf{H}_{\mu})).
\end{align*}
Furthermore, it is easily seen that the second summand of the conditional bias expression in Eq.~\eqref{Bias1}, which holds the remainder term, is given by
\begin{align*}
&\frac{1}{2}e_1^{\top}\big((nm)^{-1}[\mathbf{1},\mathbf{U}_{u},\mathbf{Z}_{z}]^{\top}\mathbf{W}_{\mu,uz}[\mathbf{1},\mathbf{U}_{u},\mathbf{Z}_{z}]\big)^{-1}\times\\
&\times(nm)^{-1}[\mathbf{1},\mathbf{U}_{u},\mathbf{Z}_{z}]^{\top}\mathbf{W}_{\mu,uz}\mathbf{R}_\mu(u,z)=o_p(tr(\mathbf{H}_\mu)).
\end{align*}\noindent
Summation of the two latter expressions yields the asymptotic approximation of the conditional bias
\begin{align*}
  \E(\hat{\mu}(u,z;\mathbf{H}_\mu)-\mu(u,z)|\mathbf{U},\mathbf{Z})=\frac{1}{2}\left(\nu_{2}(\kappa)\right)^2tr\left\{\mathbf{H}_\mu \boldsymbol{\mathcal{H}}_\mu(u,z)\right\}+o_p(tr(\mathbf{H}_\mu)).
\end{align*}

\smallskip

Proof of Lemma \ref{Bias_and_Variance_mu}, part \textit{(ii)}: In the following we derive the conditional variance of the local linear estimator $\V(\hat{\mu}(u,z;\mathbf{H}_\mu)|\mathbf{U},\mathbf{Z})=$
{\small\begin{align}
=&e_1^{\top}\big([\mathbf{1},\mathbf{U}_{u},\mathbf{Z}_{z}]^{\top}\mathbf{W}_{\mu,uz}[\mathbf{1},\mathbf{U}_{u},\mathbf{Z}_{z}]\big)^{-1}\times\notag\\
&\times[\mathbf{1},\mathbf{U}_{u},\mathbf{Z}_{z}]^{\top}\mathbf{W}_{\mu,uz}\, \mathrm{Cov}(\mathbf{Y}|\mathbf{U},\mathbf{Z})\, \mathbf{W}_{\mu,uz}[\mathbf{1},\mathbf{U}_{u},\mathbf{Z}_{z}]\times \notag\\
&\times ([\mathbf{1},\mathbf{U}_{u},\mathbf{Z}_{z}]^{\top}\mathbf{W}_{\mu,uz}[\mathbf{1},\mathbf{U}_{u},\mathbf{Z}_{z}])^{-1}e_1\notag\\\label{Varexpr}\\
=&e_1^{\top}((nm)^{-1}[\mathbf{1},\mathbf{U}_{u},\mathbf{Z}_{z}]^{\top}\mathbf{W}_{\mu,uz}[\mathbf{1},\mathbf{U}_{u},\mathbf{Z}_{z}])^{-1}\times\,\notag\\
&\times ((nm)^{-2}[\mathbf{1},\mathbf{U}_{u},\mathbf{Z}_{z}]^{\top}\mathbf{W}_{\mu,uz}\, \mathrm{Cov}(\mathbf{Y}|\mathbf{U},\mathbf{Z})\, \mathbf{W}_{\mu,uz}[\mathbf{1},\mathbf{U}_{u},\mathbf{Z}_{z}])\times \notag\\
&\times ((nm)^{-1}[\mathbf{1},\mathbf{U}_{u},\mathbf{Z}_{z}]^{\top}\mathbf{W}_{\mu,uz}[\mathbf{1},\mathbf{U}_{u},\mathbf{Z}_{z}])^{-1}e_1,\notag
\end{align}}\noindent
where $\mathrm{Cov}(\mathbf{Y}|\mathbf{U},\mathbf{Z})$ is a $nm\times nm$ matrix with typical elements
\begin{align*}
  \mathrm{Cov}(Y_{ij},Y_{\ell k}|U_{ij},U_{\ell k},Z_i,Z_\ell)=&\gamma_{|i-\ell|}((U_{ij},Z_i),(U_{\ell k},Z_\ell))+\\
  &+\sigma_\epsilon^2\mathbbm{1}{\left(i=\ell \text{ and }j=k\right)};
\end{align*}
with $\mathbbm{1}(.)$ being the indicator function.

We begin with analyzing the $3\times 3$ matrix
\begin{eqnarray*}
  (nm)^{-2}[\mathbf{1},\mathbf{U}_{u},\mathbf{Z}_{z}]^{\top}\mathbf{W}_{\mu,uz}\, \mathrm{Cov}(\mathbf{Y}|\mathbf{U},\mathbf{Z})\, \mathbf{W}_{\mu,uz}[\mathbf{1},\mathbf{U}_{u},\mathbf{Z}_{z}]
\end{eqnarray*}
using the following three Lemmas \ref{upper_left}-\ref{lower_right}.

\begin{lemma}\label{upper_left}
The upper-left scalar (block) of the matrix \\
$(nm)^{-2}[\mathbf{1},\mathbf{U}_{u},\mathbf{Z}_{z}]^{\top}\mathbf{W}_{\mu,uz}\mathrm{Cov}(\mathbf{Y}|\mathbf{U},\mathbf{Z})\mathbf{W}_{\mu,uz}[\mathbf{1},\mathbf{U}_{u},\mathbf{Z}_{z}]$ is given by
{\small\begin{eqnarray*}
  &&(nm)^{-2}\mathbf{1}^{\top}\mathbf{W}_{\mu,uz}\mathrm{Cov}(\mathbf{Y}|\mathbf{U},\mathbf{Z})\mathbf{W}_{\mu,uz}\mathbf{1}\\
  &=&(nm)^{-1}f_{UZ}(u,z)|\mathbf{H}_\mu|^{-1/2}R(K_\mu)\left(\gamma(u,u,z)+\sigma^2_\epsilon\right)(1+O_p(tr(\mathbf{H}_\mu^{1/2})))\\
  &+&n^{-1}(f_{UZ}(u,z))^2\left[\left(\frac{m-1}{m}\right)h_{\mu,Z}^{-1}R(\kappa)\frac{\gamma(u,u,z)}{f_Z(z)}+c(u,z)\right]
  (1+O_p(tr(H^{1/2})))\\
  &=&O_p((nm)^{-1}|\mathbf{H}_\mu|^{-1/2})+O_p(n^{-1}h_{\mu,Z}^{-1}),
\end{eqnarray*}}\noindent
where $c(u,z)=2\sum_{l=1}^{n-1}\gamma_{l}((u,z),(u,z))$. Under Assumption A4 there exists a constant $C$, $0<C<\infty$, such that $0\leq |c(u,z)|\leq C$.
\end{lemma}

\begin{lemma}\label{upper_right}
The $1\times 2$ dimensional upper-right block of the matrix \\
$(nm)^{-2}[\mathbf{1},\mathbf{U}_{u},\mathbf{Z}_{z}]^{\top}\mathbf{W}_{\mu,uz}\mathrm{Cov}(\mathbf{Y}|\mathbf{U},\mathbf{Z})\mathbf{W}_{\mu,uz}[\mathbf{1},\mathbf{U}_{u},\mathbf{Z}_{z}]$ is given by
{\small\begin{eqnarray*}
  &&(nm)^{-2}\mathbf{1}^{\top}\mathbf{W}_{\mu,uz}\mathrm{Cov}(\mathbf{Y}|\mathbf{U},\mathbf{Z})\mathbf{W}_{\mu,uz}\left(\begin{matrix}(U_{11}-u,Z_1-z)\\ \vdots\\ (U_{nm}-u,Z_n-z)\end{matrix}\right)\\
  &=&(nm)^{-1}f_{UZ}(u,z)|\mathbf{H}_\mu|^{-1/2}(\mathbf{1}^{\top}\mathbf{H}_\mu^{1/2})R(K_\mu)\left(\gamma(u,u,z)+\sigma^2_\epsilon\right)(1+O_p(tr(\mathbf{H}_\mu^{1/2})))\\
  &+&n^{-1}(f_{UZ}(u,z))^2(\mathbf{1}^{\top}\mathbf{H}_\mu^{1/2})\left[\left(\frac{m-1}{m}\right)h_{\mu,Z}^{-1}R(\kappa)\frac{\gamma(u,u,z)}{f_Z(z)}+c_r\right]
  (1+O_p(tr(\mathbf{H}_\mu^{1/2})))\\
  &=&O_p((nm)^{-1}|\mathbf{H}_\mu|^{-1/2}(\mathbf{1}^{\top}\mathbf{H}_\mu^{1/2}))+O_p(n^{-1}(\mathbf{1}^{\top}\mathbf{H}_\mu^{1/2})h_{\mu,Z}^{-1}),
\end{eqnarray*}}\noindent
where $c(u,z)=2\sum_{l=1}^{n-1}\gamma_{l}((u,z),(u,z))$. Under Assumption A4 there exists a constant $C$, $0<C<\infty$, such that $0\leq |c(u,z)|\leq C$.

\paragraph{Remark} The $2\times 1$ dimensional lower-left block of the matrix \\
$(nm)^{-2}[\mathbf{1},\mathbf{U}_{u},\mathbf{Z}_{z}]^{\top}\mathbf{W}_{\mu,uz}\mathrm{Cov}(\mathbf{Y}|\mathbf{U},\mathbf{Z})\mathbf{W}_{\mu,uz}[\mathbf{1},\mathbf{U}_{u},\mathbf{Z}_{z}]$ is simply the transposed version of the result in Lemma \ref{upper_right}.
\end{lemma}

\newpage

\begin{lemma}\label{lower_right}
The $2\times 2$ lower-right block of the matrix \\
$(nm)^{-2}[\mathbf{1},\mathbf{U}_{u},\mathbf{Z}_{z}]^{\top}\mathbf{W}_{\mu,uz}\mathrm{Cov}(\mathbf{Y}|\mathbf{U},\mathbf{Z})\mathbf{W}_{\mu,uz}[\mathbf{1},\mathbf{U}_{u},\mathbf{Z}_{z}]$ is given by
{\small\begin{eqnarray*}
  &&(nm)^{-2}\left(((U_{11}-u),(Z_1-z))^\top,\dots,((U_{nm}-u),(Z_n-z)^\top)\right)\times\\
  &&\times \mathbf{W}_{\mu,uz}\mathrm{Cov}(\mathbf{Y}|\mathbf{U},\mathbf{Z})\mathbf{W}_{\mu,uz}\left(\begin{matrix}(U_{11}-u,Z_1-z)\\ \vdots\\ (U_{nm}-u,Z_n-z)\end{matrix}\right)\\
  &=&(nm)^{-1}f_{UZ}(u,z)|\mathbf{H}_\mu|^{-1/2} \mathbf{H}_\mu R(K_\mu)\left(\gamma(u,u,z)+\sigma^2_\epsilon\right)(1+O_p(tr(\mathbf{H}_\mu^{1/2})))\\
  &+&n^{-1}(f_{UZ}(u,z))^2 \mathbf{H}_\mu\left[\left(\frac{m-1}{m}\right)h_{\mu,Z}^{-1}R(\kappa)\frac{\gamma(u,u,z)}{f_Z(z)}+c_r\right]
  (1+O_p(tr(\mathbf{H}_\mu^{1/2})))\\
  &=&O_p((nm)^{-1}|\mathbf{H}_\mu|^{-1/2}\mathbf{H}_\mu)+O_p(n^{-1}\mathbf{H}_\mu h_{\mu,Z}^{-1}),
\end{eqnarray*}}\noindent
where $c(u,z)=2\sum_{l=1}^{n-1}\gamma_{l}((u,z),(u,z))$. Under Assumption A4 there exists a constant $C$, $0<C<\infty$, such that $0\leq |c(u,z)|\leq C$.
\end{lemma}

Using the approximations for the bloc-elements of the matrix \\
$(nm)^{-2}[\mathbf{1},\mathbf{U}_{u},\mathbf{Z}_{z}]^{\top}\mathbf{W}_{\mu,uz}\mathrm{Cov}(\mathbf{Y}|\mathbf{U},\mathbf{Z})\mathbf{W}_{\mu,uz}[\mathbf{1},\mathbf{U}_{u},\mathbf{Z}_{z}]$, given by the Lemmas \ref{upper_left}-\ref{lower_right}, and the approximation for the matrix\\
$\left((nm)^{-1}[\mathbf{1},\mathbf{U}_{u},\mathbf{Z}_{z}]^{\top}\mathbf{W}_{\mu,uz}[\mathbf{1},\mathbf{U}_{u},\mathbf{Z}_{z}]\right)^{-1}$, given in \eqref{pre_results}, we can approximate the conditional variance of the bivariate local linear estimator, given in \eqref{Varexpr}. Some straightforward matrix algebra leads to $\V(\hat{\mu}(u,z;\mathbf{H}_\mu)|\mathbf{U},\mathbf{Z})=$
{\small
\begin{align*}
&(nm)^{-1}|\mathbf{H}_\mu|^{-1/2}\left\{\frac{R(K_\mu)\left(\gamma(u,u,z)+\sigma^2_\epsilon\right)}{f_{UZ}(u,z)}\right\}\left(1+o_p(1)\right)\\
&+n^{-1}\left[\left(\frac{m-1}{m}\right)h_{\mu,Z}^{-1}R(\kappa)\frac{\gamma(u,u,z)}{f_Z(z)}+c_r\right]\left(1+o_p(1)\right),
\end{align*}}
which is asymptotically equivalent to our variance statement of
Lemma \ref{Bias_and_Variance_mu} part \textit{(ii)}.

\noindent\textbf{Proof of Lemma \ref{upper_left}.} (The proofs of Lemmas \ref{upper_right} and \ref{lower_right} can be done correspondingly.) To show Lemma \ref{upper_left} it will be convenient to split the sum such that
$(nm)^{-2}\mathbf{1}^{\top}\mathbf{W}_{\mu,uz}\mathrm{Cov}(\mathbf{Y}|\mathbf{U},\mathbf{Z})\mathbf{W}_{\mu,uz}\mathbf{1}=s_1+s_2+s_3$. Using standard procedures from kernel density estimation leads to
{\small\begin{align}
    s_1&=(nm)^{-2}\sum_{ij}(K_{\mu,h}(U_{ij}-u,Z_{\RN{1}}-z))^2\V(Y_{ij}|\mathbf{U},\mathbf{Z})\label{M_1}\\
    &=(nm)^{-1}|\mathbf{H}_\mu|^{-1/2}f_{UZ}(u,z)R(K_\mu)\left(\gamma(u,u,z)+\sigma^2_\epsilon\right)
    +O((nm)^{-1}|\mathbf{H}_\mu|^{-1/2}\;tr(\mathbf{H}_\mu^{1/2}))\notag
\end{align}}\noindent
{\small\begin{align}    
    s_2&=(nm)^{-2}\sum_{jk}\underset{i\neq \ell}{\sum_{i\ell}}K_{\mu,h}(U_{ij}-u,Z_{\RN{1}}-z)\,\mathrm{Cov}(Y_{ij},Y_{\ell k}|\mathbf{U},\mathbf{Z})\,K_{\mu,h}(U_{\ell k}-x,Z_{\ell}-z)\label{M_2}\\
    &=n^{-1}(f_{UZ}(u,z))^2c(u,z)+O_p(n^{-1}tr(\mathbf{H}_\mu^{1/2}))\notag
\end{align}}\noindent
{\small\begin{align}     
    s_3&=(nm)^{-2}\underset{i\neq
      j}{\sum_{ij}}\sum_{t}h_{\mu,U}^{-1}\kappa(h_{\mu,U}^{-1}(U_{ij}-u))(h_{\mu,Z}^{-1}\kappa(h_{\mu,Z}^{-1}(Z_{\RN{1}}-z)))^2\,\mathrm{Cov}(Y_{ij},Y_{jt}|\mathbf{U},\mathbf{Z})\times\label{M_3}\\
    &\times h_{\mu,U}^{-1}\kappa(h_{\mu,U}^{-1}(U_{ik}-x))\notag\\
    &=n^{-1}(f_{UZ}(u,z))^2\left[\left(\frac{m-1}{m}\right)h_{\mu,Z}^{-1}R(\kappa)\frac{\gamma(u,u,z)}{f_Z(z)}\right]+O_p(n^{-1}tr(\mathbf{H}_\mu^{1/2}))\notag,
\end{align}}\noindent
where $c(u,z)=2\sum_{l=1}^{n-1}\gamma_{l}((u,z),(u,z))$. Summing up \eqref{M_1}-\eqref{M_2} leads to the result in Lemma \ref{upper_left}. Lemmas \ref{upper_right} and \ref{lower_right} differ from Lemma \ref{upper_left} only with respect to the additional factors $\mathbf{1}^{\top}\mathbf{H}_\mu^{1/2}$ and $\mathbf{H}_\mu$. These come in due to the usual substitution step for the additional data parts $(U_{ij}-u,Z_i-z)$.

\smallskip\noindent\textbf{Proofs of Theorem \ref{C_AN_mu} and Corollary \ref{cor:TSI}.} 
Theorem \ref{C_AN_mu} and Corollary \ref{cor:TSI} follow directly from Lemma \ref{Bias_and_Variance_mu} and from applying a central limit theorem for strictly stationary ergodic times series such as Theorem 9.5.5 in \citeappendix{KT75}.

\section{Data Sources}\label{sec:Data}
Hourly spot prices of the German electricity market are provided by the European Energy Power Exchange (EPEX) (\url{www.epexspot.com}), hourly values of Germany's gross electricity demand and electricity exchanges with other countries are provided by the European Network of Transmission System Operators for Electricity (\url{www.entsoe.eu}). German wind and solar power infeed data are provided by the transparency platform of the European Energy Exchange (\url{www.eex-transparency.com}). German air temperature data are available from the German Weather Service (\url{www.dwd.de}). The daily prices for natural gas are provided via the trading platform PEGAS, which is part of the European Energy Exchange (EEX) Group operated by Powernext (\url{www.powernext.com}). Daily prices for European CO$_2$ Emission Allowances (EUA) and for the Amsterdam-Rotterdam-Antwerp (ARA) coal futures are provided via the websites of the EEX (\url{www.eex.com})

\bibliographystyleappendix{imsart-nameyear}
\bibliographyappendix{bibfile_ARCHIV.bib}

\end{document}